\documentclass{article} 
\usepackage{iclr2023_conference,times}

\usepackage{amsmath,amsfonts,bm}

\def\eqref#1{equation~\ref{#1}}

\def\1{\bm{1}}

\DeclareMathAlphabet{\mathsfit}{\encodingdefault}{\sfdefault}{m}{sl}
\SetMathAlphabet{\mathsfit}{bold}{\encodingdefault}{\sfdefault}{bx}{n}

\usepackage{booktabs}
\usepackage{pifont}
\usepackage{colortbl}
\usepackage{amsmath, amssymb, amsthm}

\usepackage{multirow}
\usepackage{caption}
\usepackage{subcaption}
\usepackage{graphicx}
\usepackage{wrapfig}
\usepackage{algorithm} 
\usepackage{hyperref}
\usepackage{url}

\usepackage{algorithmicx}
\usepackage{algpseudocode}

\usepackage{amsmath}

\definecolor{lightblue}{rgb}{0.88, 0.96, 1}

\title{Jen-1 DreamStyler: Customized Musical Concept Learning via Pivotal Parameters Tuning}

\author{
Boyu~Chen\quad
Peike~Patrick~Li\quad
Yao~Yao\quad
Alex~Wang \\ 
~Futureverse, AI Innovation \\
~\texttt{\{boyu.chen, alex.wang\}@futureverse.com}
}



\iclrfinalcopy 
\begin{document}

\maketitle

\begin{abstract}
    Large models for text-to-music generation have achieved significant progress, facilitating the creation of high-quality and varied musical compositions from provided text prompts. 
However, input text prompts may not precisely capture user requirements, particularly when the objective is to generate music that embodies a specific concept derived from a designated reference collection.
In this paper, we propose a novel method for customized text-to-music generation, which can capture the concept from a two-minute reference music and generate a new piece of music conforming to the concept. 
We achieve this by fine-tuning a pretrained text-to-music model using the reference music.
However, directly fine-tuning all parameters leads to overfitting issues.
To address this problem, we propose a Pivotal Parameters Tuning method that enables the model to assimilate the new concept while preserving its original generative capabilities.
Additionally, we identify a potential concept conflict when introducing multiple concepts into the pretrained model. We present a concept enhancement strategy to distinguish multiple concepts, enabling the fine-tuned model to generate music incorporating either individual or multiple concepts simultaneously.
Since we are the first to work on the customized music generation task, we also introduce a new dataset and evaluation protocol for the new task.
Our proposed Jen1-DreamStyler outperforms several baselines in both qualitative and quantitative evaluations.
 Demos will be available at \url{https://www.jenmusic.ai/research#DreamStyler}.
\end{abstract}

\section{Introduction}
    Recent advancements in generative models~\cite{Transformer-2017-Vaswani, Improved_Diffusion-2021-Nichol, StableDiffusion-2022-Rombach} have marked significant progress in the field of text-to-music generation~\cite{MusicLM-2023-Agostinelli, MusicGen-2023-Copet, audioldm-2023-liu, jen1-2023-Li, yao2023jen}.
These models, usually trained on large-scale datasets of text-music pairs, can interpret textual descriptions to produce diverse musical compositions.
Advanced text-to-music technology allows users to experience a novel form of musical interaction, where they can input a textual description and receive a piece of music that aligns with the described mood, genre, theme, \textit{etc.}
The vastness and diversity of the training datasets enable these models to handle a wide range of musical contents (\textit{e.g., instruments}) and styles (\textit{e.g., genres}).

Despite their comprehensive training, text-to-music generation models~\cite{jen1-2023-Li} face significant challenges in fully capturing and replicating the broad spectrum of human musical concepts, which often exhibit a long-tailed distribution~\cite{Longtail-2009-Celma}.
Specifically, the models struggle with unique or context-specific musical concepts that appear infrequently and may not be included in their training datasets.
These low-frequency or novel musical concepts present substantial obstacles in the pursuit of accurate music generation.
For example, complex melodies produced by street performers using unconventional instruments, such as water cups, buckets, and chopsticks, or the unique timbre of a ventriloquist performing alone, frequently lack accurate textual descriptions. 
This highlights a notable gap in the capabilities of current models in capturing the full richness and variety of human music expression.
In light of these limitations, the pursuit of customized music generation, encompassing both content (\textit{e.g.,} unique instruments) and style (\textit{e.g.,} specific genres), becomes increasingly significant.
This highlights the vast potential and yet-to-be-realized capabilities of current text-to-music technologies.

\begin{figure*}[t]
    \centering
    \includegraphics[width=1.0\textwidth]{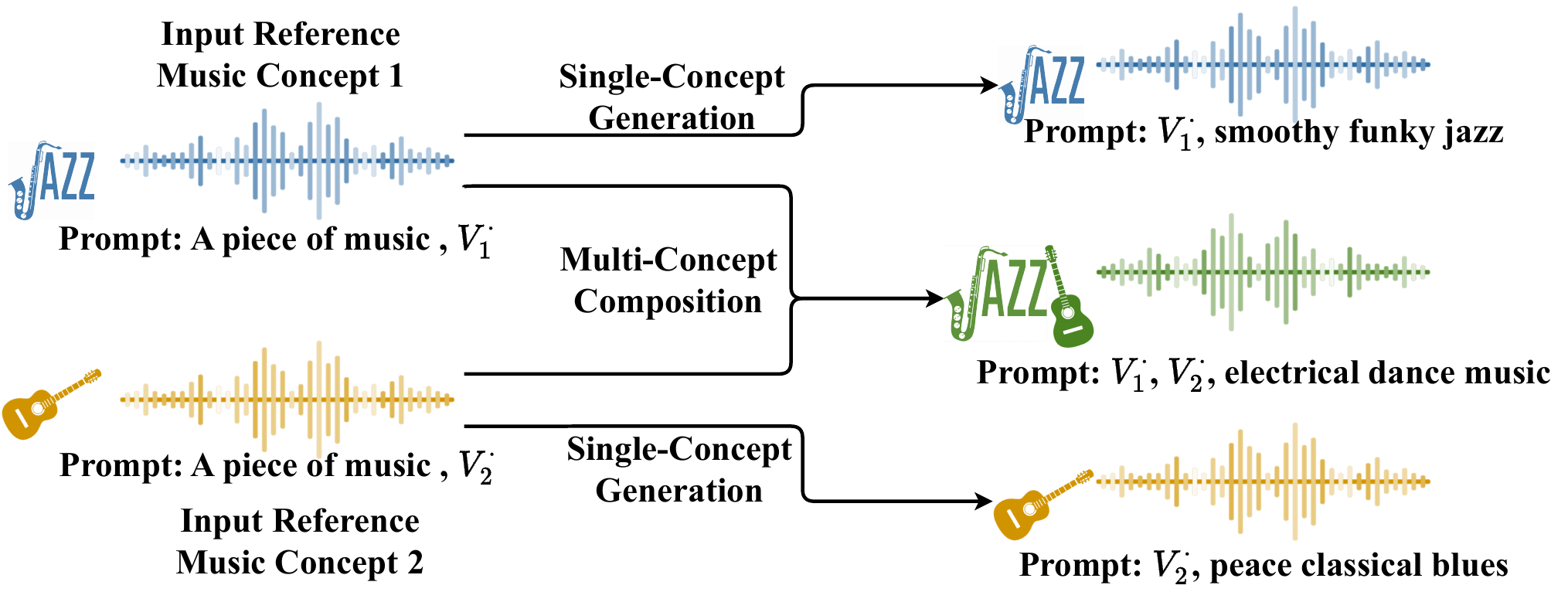}
    \caption{
    Utilizing a mere two minutes of reference music representing a new concept, our proposed JEN-1 DreamStyler can understand and reproduce the musical concept. 
    Reference musical concepts could be an instrument (\textit{e.g. guitar}), a genre (\textit{e.g.}, jazz), \textit{etc.}
    Our JEN-1 DreamStyler is not limited to mastering a single musical concept, but also proficient in simultaneously integrating and generalizing multiple musical concepts.
    }
    \label{fig:motivation}
\end{figure*}

In this work, we concentrate on customized text-to-music diffusion models by adapting them to interpret and reproduce new musical concepts, as shown in Fig. \ref{fig:motivation}.
Specifically, we aim to modifying an existing model to accurately recognize and reproduce a particular musical concept, such as an instrument or genre, without any additional textual input.
Remarkably, our approach requires only about two minutes of reference music and can operate effectively even in the absence of textual descriptions.
The primary objective is to equip the model with the capability to capture the essence of the reference music and create a range of compositions with the specific musical concept.
For this purpose, leveraging the pretrained text-to-music models for direct fine-tuning offers a straightforward approach.
Nevertheless, this method faces two significant challenges.
Firstly, there is a tendency for the model to overfit the given reference music, resulting in generated music that lacks diversity and closely resembles the reference. 
Secondly, directly fine-tuning the model to incorporate multiple musical concepts simultaneously proves to be impractical.
For instance, when attempting to merge distinct sounds from a piano and a guitar, drawn from two separate reference tracks, the model often suffers from a concept conflict issue. 
This issue typically results in one concept dominating the generation, ignoring the other, which impedes the model's capacity to effectively combine multiple musical concepts coherently.

To tackle these challenges, we propose the JEN-1 DreamStyler,  introducing an innovative regularization method named Pivotal Parameters Tuning.
This method selectively fine-tunes concept-specific pivotal parameters within the network, maintaining the remainder unchanged.
It employs a sparse mask to identify the most pivotal parameters, based on their variation relative to the reference music.
The underlying principle asserts that parameters exhibiting greater variation in mask values are pivotal for generating the target musical concept.
Consequently, these pivotal parameters are selected for subsequent fine-tuning, while the remaining parameters are kept non-trainable.
By adopting this strategy, our model effectively learns the musical concept from the reference, preserving the generality of the pretrained model to promote more diverse and generalized music with the specific concept.

Beyond the selective tuning of network parameters, our JEN-1 DreamStyler incorporates trainable identifier tokens into the input prompt.
Our goal is to improve the model’s capacity for generalization across multiple musical concepts when learned concurrently.
The conventional text-inversion method~\cite{TI-2022-Gal} typically utilizes a single extra token for each musical concept, as exemplified by phrases such as `A short piece of V$^{*}$ music'. 
However, this method is inadequate when dealing with multiple concepts,  \textit{e.g.}, `A short piece of cheerful V$_1^{*}$ and V$_2^{*}$ music'.
We observed that distinct tokens for V$_1^{*}$ and V$_2^{*}$, despite their initial uniqueness,  eventually converge into highly similar tokens after processing by the text encoder.
To resolve this issue, our model innovates by assigning multiple tokens to each musical concept.
This strategy significantly diversifies the representation of each concept within the model, ensuring that tokens corresponding to different concepts not only remain distinct but also accurately representative.
Through this enhancement, our model achieves improved generalization in capturing and distinguishing multiple musical concepts using identifier tokens.

As the initial attempt at the customized text-to-music generation task, we introduce a new benchmark dataset and an evaluation protocol.
Through a combination of qualitative and quantitative assessments, we demonstrate the effectiveness of our proposed method. 
We hope that this research will pave the way for future explorations in customized music generation,  thereby stimulating further advancements and innovations in this field.
To summarize, the contributions of this work are multi-dimensional: 
\begin{itemize}
\item \textbf{Novel Data-Efficient Framework.} We introduce an innovative framework designed specifically for data-efficient, customized music generation. This framework is capable of capturing and replicating unique musical concepts with minimal input, requiring as little as two minutes of reference music and operating effectively even without any additional textual input.
\item \textbf{Pivotal Parameters Tuning Method.} Our approach incorporates a unique, Pivotal Parameters Tuning method. 
This technique selects the pivotal parameters for generating the specific musical concept and trains only these pivotal parameters.
It focuses on learning specific musical concepts and effectively addresses the challenge of over-fitting.
\item \textbf{Multiple Musical Concept Integration.} 
We tackle the challenge of concept conflict, which occurs when multiple musical concepts are introduced simultaneously. 
Our solution employs a concept enhancement strategy that ensures each musical concept is distinctly and effectively represented within the text-to-music generation model.
\item \textbf{New Benchmark and Evaluation Protocol.} 
To support this challenging task, we have developed a novel dataset and evaluation protocol specifically tailored for customized music generation. 
This dataset serves as a benchmark for assessing our method and establishes a foundation for future research in this area.
\end{itemize}

\section{Related Work}
    \noindent\textbf{Text-to-Music Generation.}
Text-to-music generation focuses on converting textual descriptions into corresponding musical compositions.
This interdisciplinary area merges language descriptions with musical creativity, leveraging generative models to produce music that reflects the themes, moods, or tags expressed in the text.
Recent advancements in text-to-music generation have shown promising results. 
Riffusion~\cite{riffusion-2023-Forsgren}, for instance, has adapted the Stable Diffusion model for music generation. 
By converting music into mel-spectrograms, Riffusion transforms the challenging text-to-music generation into a more manageable text-to-image task. 
MusicGen~\cite{MusicGen-2023-Copet} utilizes a transformer-based autoregressive model, producing music through discrete tokens.
Its innovative delay pattern technique significantly boosts the efficiency of music generation. 
Furthermore, JEN-1~\cite{jen1-2023-Li} proposes a multi-task training framework, based on a diffusion model, that uniquely combines autoregressive and non-autoregressive training. 
This integration results in the production of high-fidelity stereo music, demonstrating the versatility and advancement in this field.
Despite these technological advancements, the field of text-to-music generation still faces substantial challenges, particularly when it comes to user interaction. One of the major challenges is the difficulty in formulating accurate and detailed text descriptions that align with user preferences. 
To address this, our work proposes a customized music generation task that does not only rely on specific text descriptions. 
Instead, our model is capable of generating various music pieces based on reference music.
This approach overcomes the challenges of text description dependency, offering a more flexible and user-friendly solution for customized music generation.

\noindent\textbf{Customized Creation using Diffusion Models.}
Customized Creation in image generation using diffusion models has become a highly popular area of research.
This approach focuses on generating images that either share a style or contain objects similar to those in reference images.
Numerous works have contributed significantly to the development of this field.
For instance, Text Inversion~\cite{TI-2022-Gal} has innovated by adding new pseudo-words to the vocabulary of a frozen text-to-image model.
This allows the model to represent a unique concept with just a single word embedding, effectively capturing a wide range of diverse and distinct ideas. 
Dreambooth~\cite{DreamBooth-2023-Ruiz} further expands on this by introducing a method to associate unique identifiers with specific subjects. 
By training the entire U-Net~\cite{U-Net-2015-Ronneberger} with their class-specific prior preservation loss, Dreambooth enables the creation of photorealistic images of these subjects in a variety of contexts and poses. 
Additionally, Custom Diffusion~\cite{CustomDF-2023-Kumari} has enhanced training efficiency by focusing on training only a portion of the parameters and utilizing regularization samples from the training dataset. They also propose a new regularization technique for multi-concept training.
Despite these advances in image generation, the concept of customization has not yet been explored in the field of music generation—until now.
Our work represents the first foray into applying the principles of customization to music generation. 
We identify and address specific challenges unique to this task and propose innovative strategies to overcome them.
Furthermore, we introduce a new dataset and an evaluation method, thus laying the groundwork for future developments in this burgeoning field.

\section{Preliminary}
    \subsection{Diffusion Model}\label{subsec:diffusionmodel}
In this work, we employ the JEN-1 model \cite{jen1-2023-Li,yao2023jen} as our foundation model, which is a state-of-the-art text-to-music generation model built upon the diffusion models. 
Diffusion models, such as those described by \cite{Denoising-2022-Ho, Improved_Diffusion-2021-Nichol}, represent an emerging class of probabilistic generative models designed to approximate complex data distributions.
These models operate by transforming simple noise distributions into intricate data representations, a process particularly effective in high-quality generation.

The diffusion model is anchored in two primary processes: forward diffusion and reverse diffusion.
In the forward diffusion phase, the model incrementally introduces Gaussian noise into the data over a series of steps. 
Each step in this Markov Chain can be mathematically expressed as
\begin{equation}
    q(x_{t}|x_{t-1}) = \mathcal{N}(x_{t}; \sqrt{1-\beta_{t}}x_{t-1}, \beta_{t}\mathbf{I}),
\end{equation}
where $x_{t}$ is the data at time step $t$ and $\beta_{t}$ are predefined noise levels.
Conversely, the reverse diffusion process involves a gradual denoising of the data. 
This is achieved through a neural network that learns to reverse the noise addition, a key element in synthesizing realistic audio. 
The reverse process can be described by the equation 
\begin{equation}
    p_\theta(x_{t-1}|x_{t}) = \mathcal{N}(x_{t-1}; \mu_\theta(x_{t}, t), \sigma^2_\theta(t)\mathbf{I}),
\end{equation}
where the functions $\mu_\theta$ and $\sigma^2_\theta$ are parameterized by the neural network, enabling the precise prediction of mean and variance at each reverse diffusion step.

The learning mechanism of diffusion models entails a fine balance between the forward diffusion process, which employs a linear Gaussian model to perturb an initial random variable until it aligns with the standard Gaussian distribution, and the reverse denoising process.
The latter utilizes a noise prediction model, parameterized by 
$\theta$, to estimate the conditional expectation 
$\mathbb{E}[\epsilon_t | x_t]$
 by minimizing a regression loss. This loss, expressed as 
\begin{equation}
    \min_\theta \mathbb{E}_{t,x,\epsilon}\left[\|\epsilon_t - \epsilon_\theta(x_t,t)\|_2^2\right],
\end{equation}
 guides the model in learning the distribution of the original data from its noisy version.

In summary, diffusion models provide a sophisticated framework for generating high-fidelity data, such as audio, by intricately modeling the transition from noise to structured data. 
This approach underlines the remarkable capability of neural networks in capturing and reproducing the complex nature of real-world phenomena.

\begin{figure*}[t]
    \centering
    \includegraphics[width=0.85\textwidth]{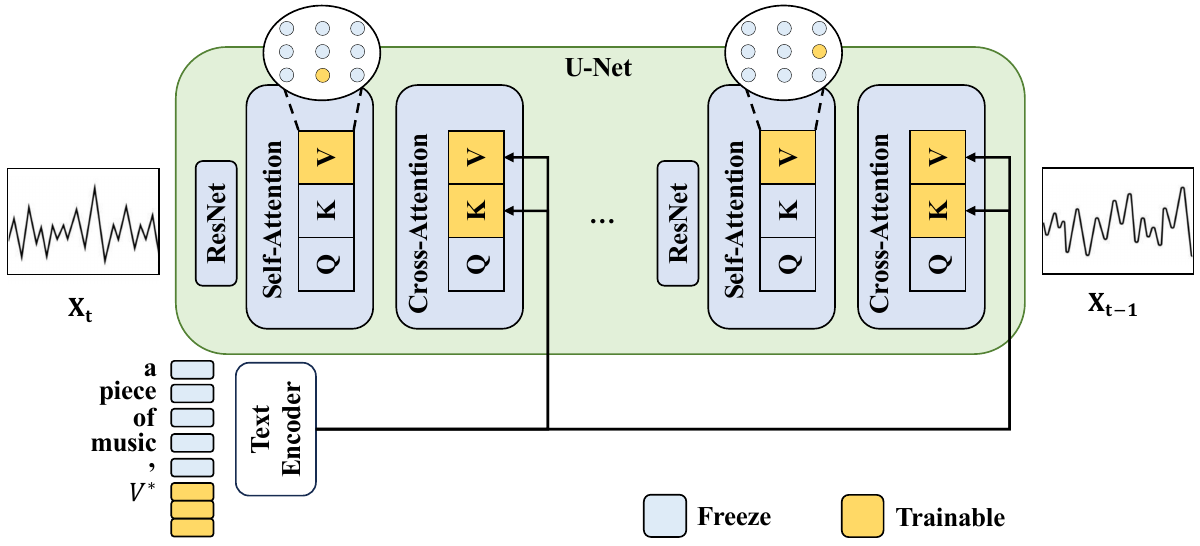}
    \caption{
    Given reference music of novel musical concepts, we select and fine-tune the most pivotal parameters within the U-Net module of our text-to-music diffusion model. 
    Furthermore, to enhance its discriminative capabilities, we introduce several trainable  concept identifier tokens, denoted as V$^*$, to present these new concepts.  
    During training, we efficiently tune these pivotal value projection parameters in the self-attention layers and all key and value projection parameters in the cross-attention layers, in conjunction with the concept identifier tokens. 
    For simplicity, we only illustrate scenarios involving the learning of a single musical concept.
    }
    \label{fig:framework}
\end{figure*}

\subsection{Text-to-Music Generation}\label{subsec:text2music}
In our method, JEN-1 serves as the foundational model for text-to-music generation, which is built based on the Latent Diffusion Model (LDM). 
This model adheres to the same forward of diffusion models mentioned in Sec \ref{subsec:diffusionmodel}, 
while the backward process and the loss function are different by incorporating textual condition $y \in \mathbb{R}^{s \times d}$ within latent space to control the synthesis process,
\begin{equation}\label{eq:text2imloss}
    \min_{\theta} \mathbb{E}_{t,x,\epsilon,y}\left[\|\epsilon_t - \epsilon_\theta(x_t,t,y)\|_2^2\right],
\end{equation}
where $x_t \in \mathbb{R}^{l \times c}$ is the noisy music latent input at timestep $t$, which is generated from the original music latent $x_0$, $\epsilon_t$ represents to stochastic noise at timestep $t$, 
and $\epsilon_\theta(\cdot)$ denotes a time-conditional 1D U-Net. 
Give the textual input features and latent music features, the textual condition $y$ is then integrated into the U-Net's intermediate layers via a cross-attention mechanism, defined as: 

\begin{equation}\label{eq:att}
\text{Attention}(Q, K, V) = \text{softmax}\left(\frac{QK^T}{\sqrt{d}}\right) \cdot V, 
\end{equation}
where,
\begin{equation}\label{eq:att}
Q = W_Q^{(i)} \cdot f^{(i)}, \quad K = W_K^{(i)} \cdot y, \quad V = W_V^{(i)} \cdot y.
\end{equation}
The matrices $W_Q^{(i)}$, $W_K^{(i)}$ and $W_V^{(i)}$ denote learnable projection parameters of the $i_{th}$ cross-attention layer. $f^{(i)} \in \mathbb{R}^{l \times c^{(i)}}$ denotes the input music feature of $i_{th}$ cross-attention layer and $y$ is the textual condition. $d$ is the output dimension of key and query features. 

The model training involves pairs of music latent and textual condition $\{(x_0, y)\}$. $\epsilon_\theta(\cdot)$ is optimized through Eq. (\ref{eq:text2imloss}). During inference, only the U-Net $\epsilon_\theta(\cdot)$is used to synthesize the desired music generation based on the textual prompt input.

In cross-attention layers within a text-to-music generation context, $W_K$ and $W_V$ project textual information, while $W_Q$ extracts music features. The attention map, computed from the interaction between music features encoded by $W_Q$ and textual features from $W_K$, is applied as weights to the textual features encoded by $W_V$. The weighted sum of textual features forms the output, enabling an effective integration of musical and textual data. 
Conversely, in self-attention layers, $W_Q$, $W_K$, and $W_V$ are all employed to encode and process the music features, facilitating internal focus on various segments of the input.

\section{Method}
    
Our proposed JEN-1 DreamStyler is designed for customized text-to-music generation, which aims to produce diverse musical compositions based on a two-minute reference piece without any supplementary textual input. 
The first challenge for the task is understanding and interpreting unique musical concepts, such as instruments or genres, associated with the reference music.
After the network has captured these musical concepts, the subsequent challenge is to produce a diverse range of music that adheres to these musical concepts.
In this section, we first introduce the new task, \textit{i.e.}, the customized text-to-music generation in Sec \ref{subsec:persona}. Then, we
show our proposed Pivotal Parameters Tuning in Sec \ref{subsec:maskpsm} to efficiently learn the musical concepts and Sec \ref{subsec:multipleconcept} to improve the quality of multiple-concept generated music.

\begin{figure*}[t]
\centering
\begin{subfigure}{.5\textwidth}
  \centering
  \includegraphics[width=0.97\linewidth]{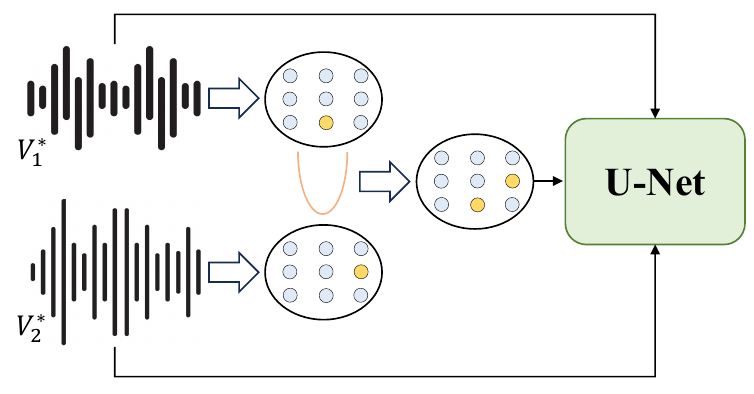}
  \caption[width=1\linewidth]{}
  \label{fig:multi_concept}
\end{subfigure}%
\begin{subfigure}{.5\textwidth}
  \centering
  \includegraphics[width=0.97\linewidth]{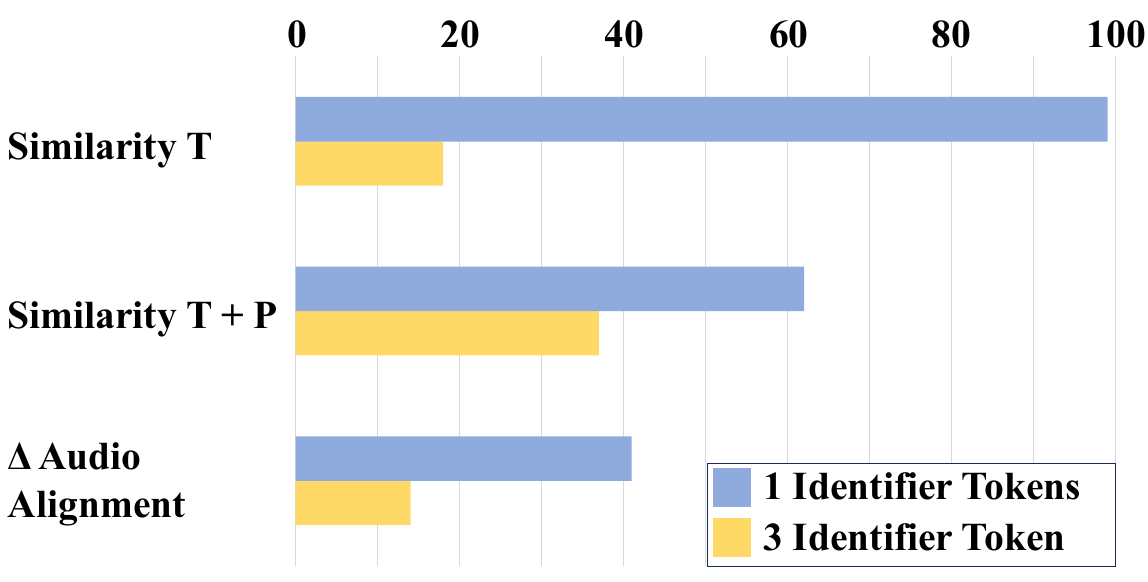}
  \caption[width=1\linewidth]{}
  \label{fig:multi_embeddings}
\end{subfigure}
\caption{Framework for multiple-concept training and comparison between different concept identifier tokens number. (a) Given two concepts, we first learn the masks for these two concepts individually and merge the two masks as the new mask for these two concepts. Then we combine the training datasets of two concepts and train the U-Net with the merged mask and dataset. We use V$_1^*$ and V$_2^*$ to represent these two concepts, respectively. 
(b) Comparison of single concept identifier token and multiple concept identifier tokens from three different aspects, including the cosine similarity between the two learned concept identifier tokens after processing through the text encoder when we only use V$_1^*$ and V$_2^*$ as input text prompt (Similarity T), or using additional rich description as `V$_1^*$, \textit{Description}' and  `V$_2^*$, \textit{Description}' (Similarity T + P).
Higher similarity means greater difficulty in distinguishing between two concepts.
Besides, we show the discrepancy of two concepts Audio Alignment Score ($\Delta$Audio Alignment), showing the distinguishing ability as in Sec \ref{subsec:benchmark}.
}
\label{fig:mutli}
\end{figure*}

\subsection{Customized Text-to-Music Generation}\label{subsec:persona} 
We propose a customized text-to-music generation method, aiming to 
understand and reproduce a new musical concept from the given reference, even without any additional textual descriptions on the concept. After integrating the new concept into the pretrained text-to-music generation model, we can utilize any text prompt to generate the music with the specific concept, such as an instrument or a given genre. The generated music will be consistent with the input text prompts, as well as the learned concept. The whole pipeline of our method is shown in Fig. \ref{fig:framework}.

With the pretrained JEN-1 model and a two-minute music clip, the intuitive approach for concept extraction is to fine-tune JEN-1 using the music clip. 
However, direct fine-tuning risks overfitting to this limited dataset, leading to a potential loss of the generalization ability. 
Regularization in neural network training effectively prevents overfitting. However, the class-specific prior preservation loss, as used in~\cite{DreamBooth-2023-Ruiz} and~\cite{CustomDF-2023-Kumari}, requires object class information, which is absent in our task.

Prior research~\cite{CustomDF-2023-Kumari}, has demonstrated the significance of cross-attention layers during the fine-tuning process and concentrated on training only the cross-attention layers, including $W_K$ and $W_V$ in Eq. (\ref{eq:att}).
Nevertheless, only training the cross-attention layers is insufficient for JEN-1 to effectively learn new concepts from the input reference music as Sec \ref{subsec:compbase} shows. 
To enhance the learning capacity of our model, we extend training to include $W_V$ from self-attention layers. 
Alongside this, we propose a Pivotal Parameters Tuning technique (details can be found in Sec \ref{subsec:maskpsm}), which facilitates an effective compromise between integrating new concepts and maintaining existing knowledge, ensuring that our model remains versatile in generating diverse musical compositions while adapting to new concepts.

\noindent\textbf{Concept Identifier Token.} 

To enhance concept extraction, we introduce a learnable concept identifier token, denoted as V$^{*}$, to represent the unique characteristics of the reference music. 
During training or generation, the concept identifier token V$^{*}$ is integrated with the original textual condition $y$ as $concat(\text{V}^{*},y)$.
Subsequently, this modification leads to an adaptation of the loss function. The original loss function, as defined in Eq. (\ref{eq:text2imloss}), is reformulated as follows:
\begin{equation}\label{eq:ourtext2imloss}
\min_{\theta, \text{V}^{*}} \mathbb{E}_{t,x,\epsilon,\text{V}^{*}}\left[\|\epsilon_t - \epsilon_\theta(x_t,t,concat(\text{V}^{*},y))\|_2^2\right].
\end{equation}

Here, the model parameters $\theta$ and the concept identifier token V$^{*}$ are trained jointly.
It should be mentioned that we may utilize more than one token to represent a new concept as in Sec \ref{subsec:multipleconcept}. For simplicity, we will still use V$^{*}$ to represent one concept in the following.

\subsection{Pivotal Parameters Tuning}\label{subsec:maskpsm}
Training only $W_K$ and $W_V$ in cross-attention layers, as done in~\cite{CustomDF-2023-Kumari}, is insufficient for our model to effectively capture concepts in reference music. 
To enhance model performance, we introduce training $W_V$ in self-attention layers as well, improving fitting ability. However, training all $W_V$ parameters can lead to overfitting, presenting a challenge in balancing concept capture and overfitting avoidance.

To tackle this issue, we introduce a Pivotal Parameters Tuning method, which selects the pivotal parameters of $W_V$ in self-attention layers for optimization. We begin by initializing a trainable mask $M_V$, which shares the same shape as $W_V$ in the self-attention block.
This mask is subsequently element-wise multiplied with $W_V$ to update it, rendering the mask $M_V$ trainable through the U-Net's forward and backward processes.
All elements in the mask $M_V$ are initialized to one, ensuring that all parameters of $W_V$ are unchanged at the beginning.
Subsequently, $M_V$ is trained using the objective,
\begin{equation}\label{eq:ourmask}
\min_{M_V} \mathbb{E}_{t,x,\epsilon,\text{V}^{*}}\left[\|\epsilon_t - \epsilon_{\{\theta, M_V\}}(x_t,t,concat(\text{V}^{*},y))\|_2^2\right],
\end{equation}
where the network parameters $\theta$ and the concept identifier token V$^{*}$ are fixed during training.

After several epochs of training the mask $M_V$, a refined mask $M_V^*$ is obtained. We then compute the mask variation as $\Delta_M = |M_V - M_V^*|$. For each parameter in $W_V$, $\Delta_M$ represent the variation of that parameter. We select the top $P\%$ of positions with the highest values in $\Delta_M$ and designate the corresponding parameters in $W_V$ as pivotal parameters, which will be optimized in the following training. These pivotal parameters, along with $W_K$ and $W_V$ from the cross-attention layers, form the trainable parameter set $\theta_{T}$. The remaining parameters are treated as non-trainable parameters, denoted $\theta_{N}$. The final training loss is defined as:
\begin{equation}\label{eq:finalloss}
\min_{\theta_{T}, \text{V}^{*}} \mathbb{E}_{t,x,\epsilon,\text{V}^{*}}\left[\|\epsilon_t - \epsilon_{\{\theta_{T}, \theta_{N}\}}(x_t,t,concat(\text{V}^{*},y))\|_2^2\right].
\end{equation}

\subsection{Multiple Concepts Integration}\label{subsec:multipleconcept}
\textbf{Joint Training on Multiple Concepts.}
As Fig~\ref{fig:multi_concept} demonstrates, to integrate multiple concepts, we first learn the mask for each concept individually and then merge the binary masks into a new mask to determine pivotal parameters for tuning.
Subsequently, we combine the training datasets for each concept and optimize pivotal parameters on the merged datasets.
To distinguish each concept, we use different concept identifier tokens to represent different concepts, i.e. V$_i^*$, and optimize them along with pivotal $W_V$ parameters in self-attention and $W_K$ and $W_V$ in cross-attention layers.
The entire pipeline is illustrated in Algorithm \ref{algo:multi}.

\begin{algorithm}[t]  
\caption{Pipeline for Multiple Concepts Integration}
\label{algo:multi}
\flushleft \textbf{Input:} \\
Reference Music Clips \{\(x^1\), \(x^2\), $\dots$, \(x^n\) \}, \\
Pretrained U-Net $\epsilon_{\theta}(\cdot)$ \\
\flushleft \textbf{Process:} \\
\begin{algorithmic}[1]  
\For{$i \in \{0, 1, \hdots, n\}$}
\State Initialize concept tokens V$_i^{*}$ and trainable mask $M^{i}$
\State Train the mask $M^{i}$ on \(x^i\) via Eq. (\ref{eq:ourmask})
\State Get the set of pivotal parameters $\theta_T^i$ according to $M^i$
\EndFor
\State Take the union of $\{\theta_T^i\}_{i=1}^{n}$ to get the pivotal parameters $\theta_T$
\State Optimize $\{V_i^*\}_{i=1}^{n}$ and $\theta_T$ on whole dataset \{\(x^1\), \(x^2\), $\dots$, \(x^n\) \} via Eq. (\ref{eq:finalloss})
\end{algorithmic}
\textbf{Output:} Optimized $\{V_i^*\}_{i=1}^{n}$ and $\theta_T$.
\end{algorithm}

\noindent\textbf{Concept Enhancement Strategy.}
In joint training involving multiple concepts, it is essential that the learned concept identifier tokens, denoted as V$_i^*$ for different concepts, are distinct from each other. However, our observations indicate that a single concept identifier token for each concept often turn to similar after processing through the text encoder.
Fig \ref{fig:multi_embeddings} compares the outcomes of using one concept identifier token versus multiple concept identifier tokens for each concept. For simplicity, this discussion focuses on just two concepts.

Initially, we examine the cosine similarity of two learned concept identifier tokens after processing through the text encoder when only V$_1^*$ and V$_2^*$ are utilized as text prompt for generation. 
This approach results in a similarity exceeding 99\%, rendering it challenging to differentiate between the two concepts under these conditions. 
To address this limitation, we augment the input text prompts with more muscial description, changing it to `V$_1^*$, \textit{Description}' and  `V$_2^*$, \textit{Description}'. This modification reduces the similarity score, but it is still above 60\%.

These similarity scores are indicative of the discriminative capacity of the concept identifier tokens, a crucial factor for generating optimal music that incorporates multiple concepts. When the similarity score is high, V$_1^*$ and V$_2^*$ are likely to converge on the same concept, leading the model to generate music that predominantly reflects one concept while neglecting the other. 
The $\Delta$Audio Alignment Score (details can be found in Sec \ref{subsec:ablation}) further substantiates this, showing a significant discrepancy in Audio Alignment Scores between the two concepts when only a single concept identifier token is used for each concept. Higher $\Delta$Audio Alignment indicates the model is more likely to generate only one concept rather than the simultaneous generation of the two concepts as we expect.

Based on this experiment, we increase the concept identifier tokens number for each concept, according to the following reasons: 
(1) Richer Representation: More tokens per concept lead to a richer, more distinct representation, reducing the risk of similarity for different concepts.
(2) Minimized Overlap: Increasing the number of tokens helps decrease overlap in the conceptual space, especially important for closely related concepts.
(3) Adaptive Flexibility: A higher count of tokens allows the model to better adapt to the complexities and variations of musical concepts, enhancing its ability to differentiate subtle nuances.
This concept enhancement strategy significantly improves the model's discriminative ability for multiple concepts, ensuring a more accurate representation in complex musical compositions.
Applying the proposed strategy leads to a reduction in all key metrics presented in Fig \ref{fig:multi_embeddings}. This decline in metrics is indicative of the enhanced discriminative ability of our model when handling multiple concepts.

\section{Experiment}
    
In this paper, we propose a new task of customized music generation. To facilitate this, we establish a new benchmark, detailed in Sec \ref{subsec:benchmark}, which includes both the Dataset and the Evaluation Protocol. Subsequent Sec \ref{subsec:impdetails} shows the implementation details of our experimental approach. Then, we present a comparative analysis of our method against a selection of baseline models to highlight its efficacy in Sec \ref{subsec:compbase}. Finally, the paper concludes with an in-depth ablation study in Sec \ref{subsec:ablation}, providing insights into the contributory elements of our method.

\subsection{Dataset and Evaluation}\label{subsec:benchmark}
\noindent\textbf{Dataset.}
We collected a benchmark
of 20 distinct concepts, including a balanced collection of 10 musical instruments and 10 genres, such as Erhu, Kora, Muzak, Urban, \textit{etc.} 
The audio samples for this dataset were sourced from various online platforms.
For each concept, we collect a two-minute audio segment to form the training set, supplemented by an additional one-minute audio segment that serves as the evaluation set.
Further enriching our dataset, we also collected 20 prompts from MusicCap~\cite{MusicCap-2021-Manco} dataset, which were specifically chosen for their diversity in content and style. These prompts were utilized to evaluate the versatility and robustness across various musical themes. 
The full list of prompts can be found in the supplementary materials. 
In our evaluation suite, we generated 50 audio clips for each concept and prompt, resulting in a total of 20,000 clips. This extensive compilation enables a thorough assessment of method performance and generalization capabilities.
We will make both the dataset and evaluation protocol available to the public via the project webpage, to facilitate future research in subject-driven audio generation.

\begin{table*}[!t]
\centering
\caption{Quantitative comparisons. Our method achieves the best two-type alignment balance.}
\label{table:baseline}
\resizebox{\textwidth}{!}{
\begin{tabular}{l|c|c|c|c}
\toprule
 & Tuned & Text & Audio & Preference \\ 
 & Parameters & Alignment $\uparrow$ & Alignment $\uparrow$ & Ratio $\uparrow$ \\ 
\midrule
Train Identifier Token Only & 0.001M & 34.70 & 27.41 & 6.4 \\ 
Train All Parameters in U-Net & 746.02M & 15.89 & 61.65 & 9.8 \\ 
Train Cross-Attn KV \& Identifier Token & 25.56M & 26.60 & 23.30 & 11.5 \\ 
\midrule
JEN-1 DreamStyler-Single & 26.18M & 29.39 & 37.07 & 72.3 \\ 
JEN-1 DreamStyler-Multiple & 26.81M & 22.24 & 44.73 & $\mathbf{-}$ \\ 
\bottomrule
\end{tabular}
}
\end{table*}

\noindent\textbf{Evaluation Metrics.}
We evaluate our method based on three metrics, the first two of which are similar to those proposed in Textual Inversion~\cite{TI-2022-Gal}. 

(A) \textbf{Audio Alignment Score}, which measures the similarity between the generated audio and the target concept. 
It shows the model's ability to learn new concepts from the reference music.
Specifically, the CLAP~\cite{CLAP-2023-Elizalde} model is utilized to calculate the CLAP space features. The cosine similarity between features from the generated audio and the target concept is calculated to determine the Audio Alignment Score.
In the context of multi-concept generation, the audio alignment for each target concept within the generated audio is computed separately. The mean of these values is then taken as the final  Audio Alignment Score. 

(B) \textbf{Text Alignment Score}, which evaluates the ability of methods to generate target concepts that are aligned with corresponding textual prompts. 
For this purpose, we generate audio segments using a diverse array of prompts, varying in content, style, and theme. Subsequently, we computed the average CLAP-space feature of these generated audio segments. The Text Alignment Score is then determined by calculating the cosine similarity between this average CLAP-space feature and the CLAP-space features of the textual prompts without the concept identifier token V$^*$.

(C) \textbf{$\Delta$Audio Alignment score}, which is utilized only in the context of multiple-concept learning, to evaluate the model tendency. 
In the multiple-concept learning, the $\Delta$Audio Alignment score is the discrepancy between the Audio Alignment Score for each target concept.
Higher $\Delta$Audio Alignment indicates the model is more likely to generate only one concept rather than the simultaneous generation of the two concepts as we expect.
Our ultimate objective is to distinctly learn different concepts for multiple concepts. Therefore, a model achieving a lower $\Delta$Audio Alignment score is considered more effective in this regard.

Audio Alignment Score and Text Alignment Score are used in both single-concept learning and multiple-concept learning. While $\Delta$Audio Alignment score is only used in multiple-concept learning.

\subsection{Implement details}\label{subsec:impdetails}
We utilize JEN-1~\cite{jen1-2023-Li} model as the pretrained model. The textual condition features are extracted by FLAN-T5~\cite{T5-2022-Chung} before sending into the U-Net model.
All experiments are conducted using an A6000 GPU and Pytorch framework. 
Before network training, we initially dedicate 100 epochs to training the mask for Pivotal Parameters selection.
For the training process, we configure the model with a batch size of 32, a learning rate of 1e-5 for U-Net parameters and 1e-4 for learnable concept identifier tokens, respectively. We adopt $\beta_1=0.9$, $\beta_2=0.95$, a decoupled weight decay of 0.1, and gradient clipping of 1.0.
We train the model for 1,500 steps with AdamW optimizer~\cite{AdamW-2010-Loshchilov} for both single and multiple concepts. 
The number of concept identifier token is set to 3 without further declaration.
For a fair comparison, we use 200 steps of classifier-free guidance~\cite{CLassifierFree-2022-Ho} with a scale of 7 for all experiments during the music generation.

\begin{table}[b]
\centering
\caption{Ablation study on Training Parameter Ratio and Parameter Selection.}
\label{table:parameter}
\begin{tabular}{c|c|c}
\toprule
Training  Ratio (\%)& Text Alignment $\uparrow$ & Audio Alignment $\uparrow$\\ 

 \midrule
1                              &     29.39        &     37.06         \\ 
\textbf{5}                              &     \textbf{26.01}        &     \textbf{39.91}         \\ 

10                             &     24.11        &     42.23         \\ 
50                             &     19.43        &     46.10         \\ 
100                            &     18.67        &     46.68         \\ 
\midrule
5-random                       &     28.14        &     35.64         \\ 
\bottomrule
\end{tabular}
\end{table}

\subsection{Comparisons with baseline}\label{subsec:compbase}
In our approach, we train three distinct sets of parameters: (1) all key and value projection parameters of cross-attention layers, (2) pivotal value projection parameters of self-attention layers, and (3) the learnable concept identifier token for new concept.
Building on this, we generate three baseline models for comparative analysis. The first baseline optimizes solely the learnable concept identifier tokens for new concepts, consistent with the methods used in ~\cite{TI-2022-Gal}. The second baseline model diverges by keeping the tokens for new concepts fixed while fine-tuning all parameters in the diffusion model. Here, each target concept is represented by a unique identifier, e.g., `sks', a token infrequently used in the text token space and not adjusted during fine-tuning as in~\cite{DreamBooth-2023-Ruiz}. In the third baseline, we limit fine-tuning the key and value projection parameters in the cross-attention layers of the U-Net, introducing a new V$^{*}$ token for the new concept while keeping other parameters fixed, as in~\cite{CustomDF-2023-Kumari}.

As demonstrated in Table~\ref{table:baseline}, our method outperforms these baselines considering the balance of Text and Audio Alignment. 
Our approach's superiority over the first baseline can be attributed to the training of a broader variety of parameters, enhancing the model's ability to extract new concepts from the reference music. In contrast, training that focuses solely on concept identifier token proves insufficient for learning concepts from reference music. While such training might yield a higher Text Alignment Score, it often results in generated music that scarcely reflects the concept of the reference. This discrepancy leads to suboptimal results in the Audio Alignment Score.

\begin{table*}[t]
\centering
\caption{Ablation study on Concept Identifier Token Number for single and multiple concepts. $\Delta$Audio Alignment is the difference between the Audio Alignment Score of two concepts for multiple-concept learning.}
\label{table:text embedding number}
\setlength{\tabcolsep}{6mm}{
\begin{tabular}{c|ccc}
\toprule
\multicolumn{1}{c|}{\multirow{2}{*}{}} & \multicolumn{3}{c}{Concept Identifier Tokens Number}             \\ \cline{2-4} 
\multicolumn{1}{c|}{}                  & \multicolumn{1}{c|}{1}    & \multicolumn{1}{c|}{3}     & \multicolumn{1}{c}{5}      \\ 
\midrule
Text Alignment-Single $\uparrow$   & \multicolumn{1}{c|}{25.87} & \multicolumn{1}{c|}{26.17} & \multicolumn{1}{c}{26.01} \\ 
Audio Alignment-Single $\uparrow$ & \multicolumn{1}{c|}{38.24} & \multicolumn{1}{c|}{37.33} & \multicolumn{1}{c}{39.91} \\ 
\midrule
Text Alignment-Multiple $\uparrow$ & \multicolumn{1}{c|}{21.99} & \multicolumn{1}{c|}{22.25} & \multicolumn{1}{c}{17.63} \\ 
Audio Alignment-Multiple $\uparrow$ & \multicolumn{1}{c|}{42.55} & \multicolumn{1}{c|}{44.73} & \multicolumn{1}{c}{44.43} \\ 
$\Delta$Audio Alignment $\downarrow$  & \multicolumn{1}{c|}{24.38} & \multicolumn{1}{c|}{8.05}  & \multicolumn{1}{c}{12.20} \\
\bottomrule
\end{tabular}}
\end{table*}

While the second model trains more parameters than ours, it still underperforms, illustrating that the generation ability of a model depends not only on the quantity but also on the type of trained parameters. Specifically, training all parameters in the U-Net model can lead to substantial overfitting to the reference music, making the text prompt losing the ability to control the generation. As shown in Table~\ref{table:baseline}, Training All Parameters in U-Net gets the lowest score in Text Alignment.

The third baseline, although it incorporates learnable concept identifier tokens and partial network parameter training, falls short of our model's performance. Training only KV in cross-attention layers is not enough to learn the concept from the reference music, leading to poor performance on Audio Alignment.
This highlights the necessity of carefully balancing the number of trainable parameters to effectively learn new concepts without losing the prior knowledge of the pretrained model.

For qualitative evaluations, we employ a Preference Ratio derived from human evaluations to assess the quality of customized generation by various methods.
Specifically, we collect 100 unique samples from each method (resulting from 10 prompts multiplied by 10 music references), leading to 400 music samples in total. 
Every sample from the set of 100 unique samples features a distinct combination of prompt and music reference. 
Specifically, we structured 100 tasks, each listing four music samples generated from the different methods alongside the corresponding prompt and reference music. Each rater performed 100 tasks, selecting their preferred sample from sets of four (one from each method) based on text alignment, audio alignment, and overall music quality.
The Preference Ratio of each method was computed by dividing the number of selected samples by the total number of tasks and is expressed as a percentage. 
A higher Preference Ratio indicates a stronger preference for a particular method.
The results, presented in Table~\ref{table:baseline}, demonstrate a significant preference for our method, indicating its superior ability to generate high-quality music that effectively meets the criteria, thus setting a promising standard in the emerging field of customized music generation.

\subsection{Ablation Studies}\label{subsec:ablation}
In this section, we conduct experiments to understand how different components affect the performance of our model. We focus on the Pivotal Parameters selection and examine two key areas. First, we look at how the ratio of training parameters influences the final results. Then, we compare our selection method with random selection to show its effectiveness. For the integration of multiple concepts, we also investigate the effect of using different numbers of concept identifier tokens.

\textbf{Training Parameter Ratio.} 
In the Pivotal Parameters Tuning approach, we selectively train a subset of influential value projection parameters from the self-attention layers. The selection ratio is varied from 1\% to 100\%, as detailed in Table~\ref{table:parameter}. 
Increasing the ratio will improve the Audio Alignment ability but hurt the generalization ability of our model.
Our results indicate that a selection ratio of 5\% yields optimal performance. At this ratio, the model effectively balances the acquisition of new concepts with the preservation of previously learned knowledge.

\textbf{Compared with Random Selection.} 
Our study also includes a comparison between our Pivotal Parameters and random selection. 
As shown in Table~\ref{table:parameter},
the comparison between `5' and `5-random' shows that training the parameters chosen through our Pivotal Parameters method brings the model superior fitting capabilities and results in a better Audio Alignment compared to training those selected randomly.

\textbf{Concept Identifier Token Number.} 
In Table~\ref{table:text embedding number}, we present the model's performance in terms of text and audio alignment with varying numbers of concept identifier tokens. 
In the context of Single Concept learning, variations in the number of concept identifier Tokens show minimal impact on performance.
However, in multiple-concept learning (we use two concepts here), despite similar Text and Audio Alignment when using either 1 or 3 concept identifier tokens, the $\Delta$Audio Alignment of using 1 concept identifier token is much higher than that of using 3 concept identifier tokens. 
This suggests a strong bias toward one of the concepts, which is contrary to our expectations for multiple-concept learning. Consequently, we have opted for using 3 concept identifier tokens in our approach to ensure a balance between distinct concept learning and computational efficiency.

\section{Conclusion}
    In this paper, we introduce a new customized music generation task and a corresponding framework for this task. We utilize learnable concept identifier tokens to represent new concepts and fine-tune the large-scale text-to-music diffusion model using just a two-minute reference track. To balance the trade-off between learning new concepts while maintaining prior knowledge, we introduce a Pivotal Parameters Tuning method and optimize only the selected parameters in the diffusion model. To address the conflicting issues when introducing multiple concepts during music generation, we present a concept enhancement strategy, which greatly improves the quality of generated music featuring multiple concepts. Furthermore, we have established a benchmark and developed evaluation protocols for this customized music generation task. We anticipate that this benchmark will facilitate future research on this topic.



\bibliography{iclr2023_conference}
\bibliographystyle{iclr2023_conference}


\end{document}